\documentclass[twocolumn]{aastex62}
\usepackage{graphicx}
\usepackage{amsmath}

\providecommand{\xmm}{{\it XMM-Newton}}
\providecommand{\cha}{{\it Chandra}}
\providecommand{\dd}{\mathrm{d}}

\renewcommand{\arraystretch}{1.5}
\newcommand{\obj}{NGC\,4051}

\begin{abstract}
	\obj\ is one of the X-ray brightest and most variable Seyfert galaxies.
	During 2009 \obj\ was caught at its lowest state ever observed by \xmm.
	The low source continuum emission allows a clear measurement of the He-like emission lines of N$^{+5}$ and O$^{+6}$. 
	The exceptionally high intercombination line flux is a direct measurement of either a high density or a UV irradiated photoionized plasma. Either way, this provides an unambiguous distance diagnostic.
	We find that the line emitting region has a density of $\sim3\times10^{10}$ cm$^{-3}$ and distance of a few $10^{15}$ cm from the continuum source, placing it within the UV-optical Broad Line Region of \obj. 
	Both N$^{+5}$ and  O$^{+6}$ independently result in similar measurements for density and distance. 
	In addition, we find the kinematics of the broad O$^{+7}$ Ly$\alpha$ match those of the He$^{+1}\,\lambda4686$\,\AA\ line, which is associated with the \obj\ Broad Line Region. This is the first direct X-ray measurement of a Seyfert Broad Line Region.
\end{abstract}

\begin{document}
	\author{Uria Peretz}
	\affiliation{Department of Physics, Technion, Haifa, Israel}
	\author{Jon M. Miller}
	\affiliation{Departemnt of Astronomy and Astrophysics, University of Michigan 48109-1107, Michigan, USA}
	\author{Ehud Behar}
	\affiliation{Department of Physics, Technion, Haifa 3200003, Israel}
	
	\title{Direct observation of the Broad Line Region in X-rays during the low state of \obj}

	\section{Introduction}
	Optical and UV emission lines with velocity broadening are ubiquitously observed in spectra of Active Galactic Nuclei (AGN). This broadening is attributed to a virial Keplerian velocity around a black hole (BH), $M_B\sim V^2 r/G$. When $V\gtrsim1000$ km $s^{-1}$ the emitting region at distances $R_\mathrm{BLR}$ has been termed the Broad Line Region. The spectral lines are commonly seen in lower ionization species such as H$\alpha$, H$\beta$, He$^{+1}$, and C$^{+3}$ as broad spectral features. These lines often react to changes in the UV and optical continuum within a few light days for low mass Seyferts, placing the BLR at $\sim10^{15}-10^{16}$ cm from the continuum source.
	
	While the BLR is commonly observed in the longer wavelengths, in X-rays it has not been measured directly. The BLR was known to absorb X-rays, initially serving as a probe for BLR parameters \citep[e.g.][]{Reichert86}. With the advent of the X-ray spectrometers on board \xmm\ and \cha\ observations of broad emission features were made, such as \citet{Kaastra02} and \citet{Blustin09}, though these are always measured tentatively in the residuals, with the distances only inferred through line width estimates and velocities weakly constrained.
	Another possibility of observing the BLR in X-rays is through obscuration of the X-ray source \citep{Risaliti11a,Risaliti11b}, where the timescale of obscuration is translated into the $R_\mathrm{BLR}$. \citet{Svoboda15} associated the variability timescale of ionized absorbers with the BLR, though it is unclear if these are just unrelated outflows at similar distances or an intrinsic part of the UV-optical BLR.
	
	A more complex analysis is the \textit{"locally optimally emitting clouds"} \citep{Badlwin95} model. \citet{Costantini07} find that in Mrk 279 this estimate is enough to account for the observed X-ray line luminosities, though results were more ambiguous in \citet{Costantini16}.
	
	Finally, \citet{Miller18} observed line flux variations in the Fe K$\alpha$ line of NGC 4151 on time scales of $\sim10^4$ s. This places the emitting region at $500-1000$ $GM_\mathrm{BH} / c^2$, within the inner BLR or X-ray BLR, though this may not be what is commonly referred to as the UV-optical BLR, which is considered to be further out. These observations may reveal that the BLR extends more towards the inner part of the accretion system than previously thought.
	
	In this paper we present the first direct diagnostic of an AGN BLR seen in X-ray emission lines, using the O$^{+6}$ and N$^{+5}$ He-like lines, considering both effects of collisional excitation and UV pumping on the line intensity ratios. In addition we compare  the broad spectral O$^{+7}$ line to other broad lines in the optical and UV spectra of \obj.
	
	\section{\obj\ and previous works}\label{sec:prev}
	\obj\ is a nearby Seyfert 1 at redshift $z=0.00216$, distance $D_L=9.3$ Mpc, mass of $M_{BH}=2\times10^6 \,M_\odot$,  with a bolometric luminosity of $L_\mathrm{bol}=3.46\times10^{42}$ erg s$^{-1}$. 
	In addition it is highly variable in X-rays, with the flux changing drastically on timescales of minutes (See Fig. \ref{fig:lc}). As such it has been observed with the Reflection Grating Spectrometer (RGS) on-board \xmm\ 17 times, 3 times with the \cha\ transmission gratings, and even more with RXTE and Swift. The broadband spectral variability of \obj\ has been extensively discussed by \citet{Uttley04}, who correlate different spectral components in an \xmm\ observation from 2002.
    
    The physics of \obj\ has been examined by X-ray gratings several times. First were \citet{Collinge01} who measured several outflowing ionized components, comparing to UV spectra and showing the most ionized X-ray components are not seen in the UV. \citet{Mason04} find evidence for a broad, possibly relativistic O line. 
    
    \citet{Lobban11} examined \cha\ transmission grating spectra and found several outflowing ionization components, with possible significant feedback to the host galaxy. More recently, \citet{Silva16} modeled NGC 4051, and measured timing lags consistent with ionized absorber distances of a few $10^{15}$ cm.
	
	The present paper is not the first to examine the possibility of the BLR in X-rays. \citet{Steenbrugge09} examined this possibility by comparing the broad emission wings of O$^{+6}$ and He$^{+1}$ from the UV spectrum of \citet{Peterson00}, finding consistent velocities and concluding, as did \citet{Costantini07}, that the BLR is stratified. 
	
	The O$^{+6}$ emission line fluxes have been used previously as a density diagnostic for \obj. 
	\citet{Pounds04} measure the forbidden to intercombination ratio ($f/i$) in the same observation discussed by \citet{Uttley04}, and find the gas to be in the low electron density limit. They estimate the size of the X-ray line emitting region to be $\sim3\times10^{17}$ cm, more extended than where the BLR is observed in the UV or optical \citep{Peterson00,Shemmer03,Fausnaugh17}. \citet{Ogle04} observing in yet another epoch also try and measure the $f/i$ ratio in their observation and also find the density to be below the critical density. They infer a possible observation of the BLR in X-rays from the broadening in the O and C lines.
   
	\section{Observations}\label{sec:obs}
    
    \begin{table}
        \centering
        \caption{Low state observations.}\label{table:obs}
        \begin{tabular}{c|ccc}
            Observation & Date       & Exposure     & Rate \\	
            &            & s     & Counts s$^{-1}$ \\	
            \hline	
            0606320401  & 2009-05-11 & 28800& $0.141\pm0.002$ \\
            0606321901  & 2009-06-02 & 44500 & $0.167\pm0.001$ \\
            0606322101  & 2009-06-08 & 37650 & $0.106\pm0.001$ \\
        \end{tabular}
    \end{table}
	
	While \obj\ has a plethora of high resolution X-ray grating observations, we are primarily interested in the lowest states of \obj\ as this is when the emission lines are most prominent and absorption contamination is least. In Table \ref{table:obs} the three lowest-state \xmm\ observations are listed along with their exposures and count rates. We limit the low state to be less than 0.2 counts s$^{-1}$ in the RGS. The 2002 observation previously analyzed by \citet{Uttley04} and \citet{Pounds04} has 0.220 counts s$^{-1}$. 
		
\begin{figure}
    	\includegraphics[trim=1cm 0 1cm 1cm,clip,width=\linewidth]{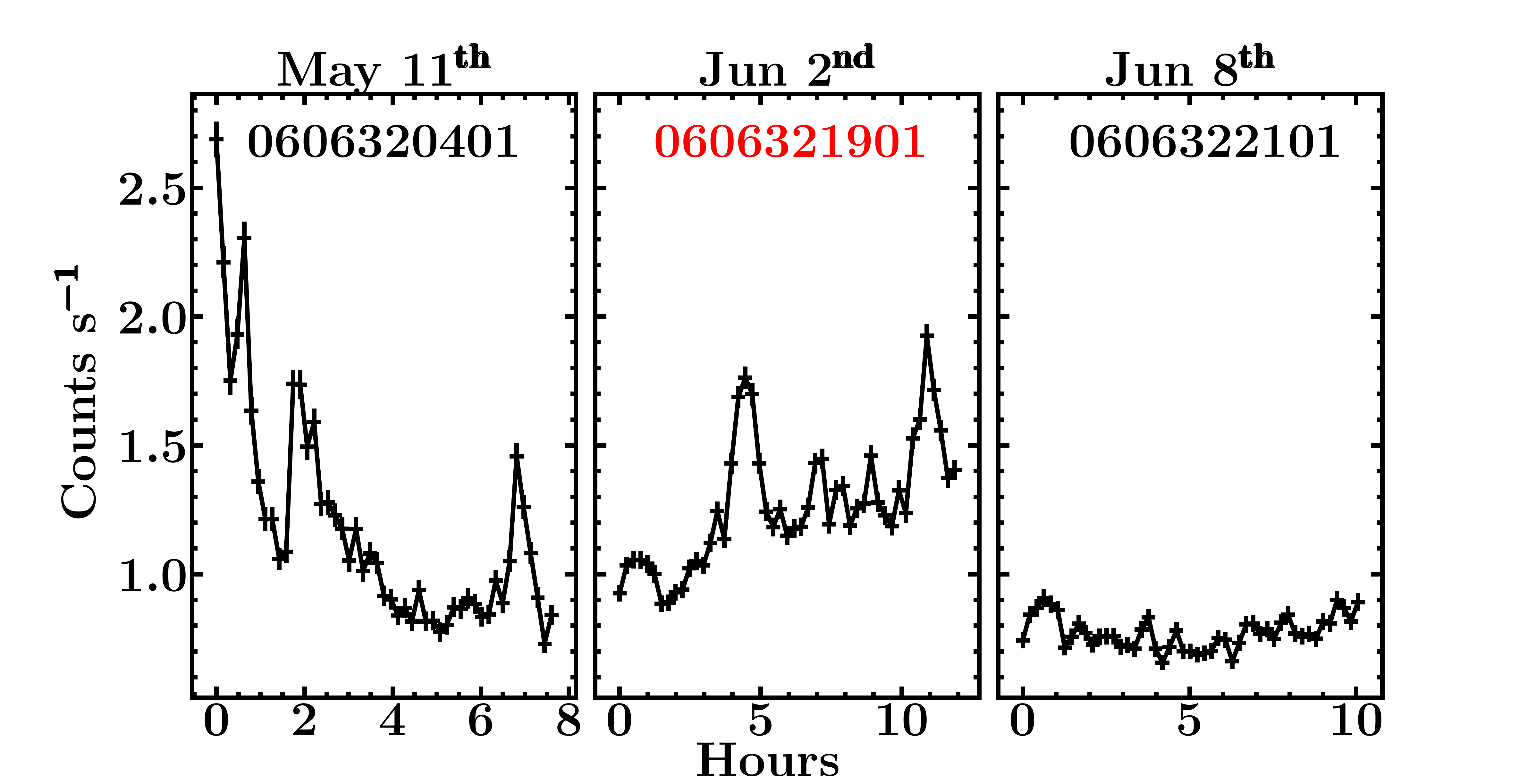}
        \caption{The 2009 M1 lightcurve of the observations discussed in this paper. The source flux changes in an order of minutes. The best diagnostics are obtained from observation 0606321901, in red.}\label{fig:lc}
    \end{figure}

	Stacking spectra is a common practice in X-ray spectroscopy, and in particular for low flux states such as those we use in this paper. 
    In Fig. \ref{fig:lc} we show the \xmm\ MOS1 (M1) light curves. It is evident that \obj\ is variable on timescales of less than an hour, and the benefit of added statistics may be completely annulled by time dependence of the diagnostic features. 
    In the observations listed in Table \ref{table:obs}, the broad wings of the intercombination and forbidden line change between the observations, and summing them creates a skewed spectral feature that impedes the narrow line flux measurement.
    As such, we perform our measurements on each of the observations individually.
    
    We note the RGS spectra themselves already introduce an ambiguity as they are a result of a $\sim10$ hour exposure. Depending on how fast the emission features react to changes in continuum flux, this uncertainty may or may not be significant. A single good measurement of $f/i$ requires at least 10 hours of exposure with the \xmm\ RGS.
    
%
    
    \section{The He-like triplets}\label{sec:lines}
    
    \begin{figure}
        \includegraphics[trim=0cm 0 3cm 2.5cm,clip,width=\linewidth]{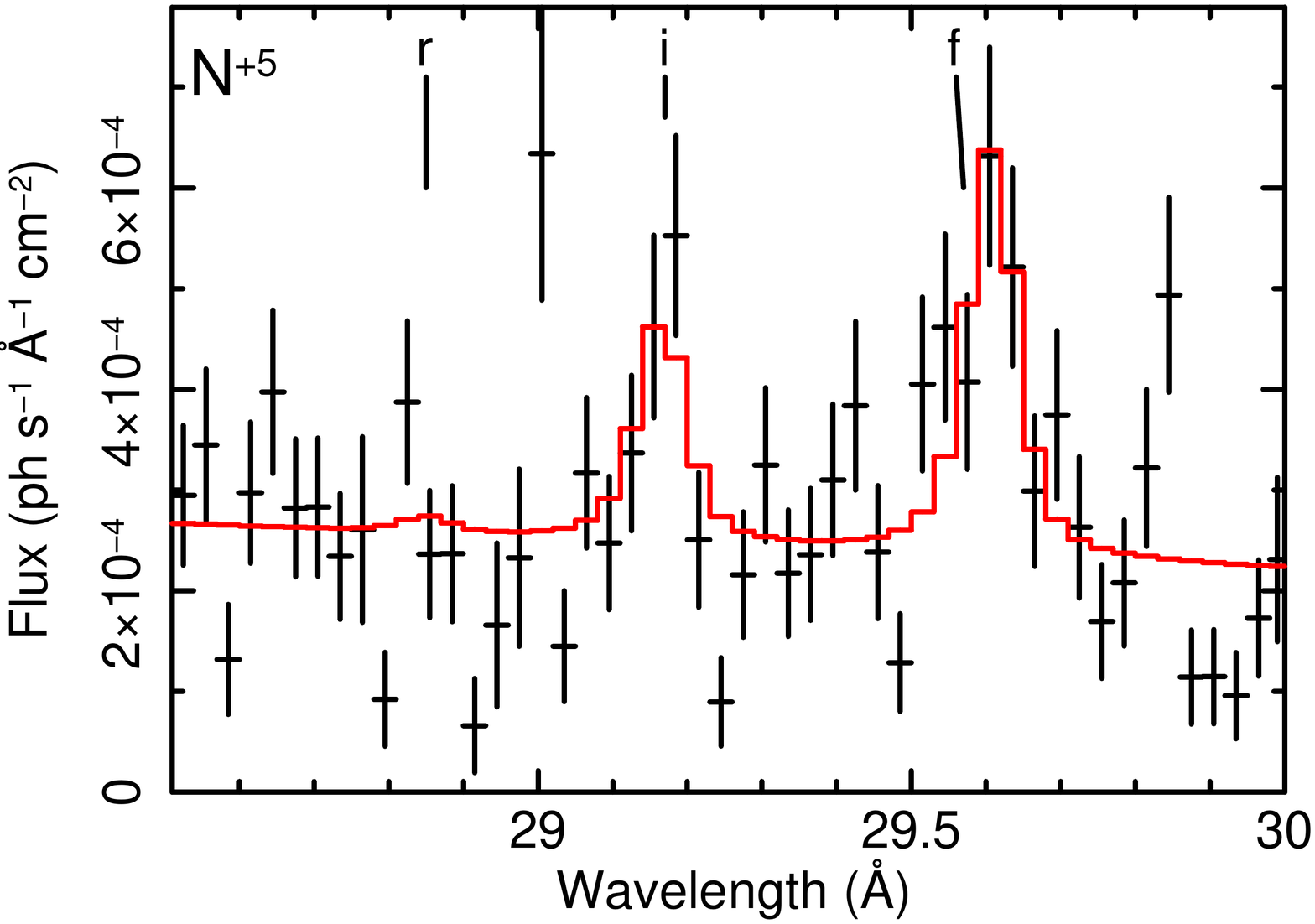}\\
        \includegraphics[trim=0cm 0 3cm 2.5cm,clip,width=\linewidth]{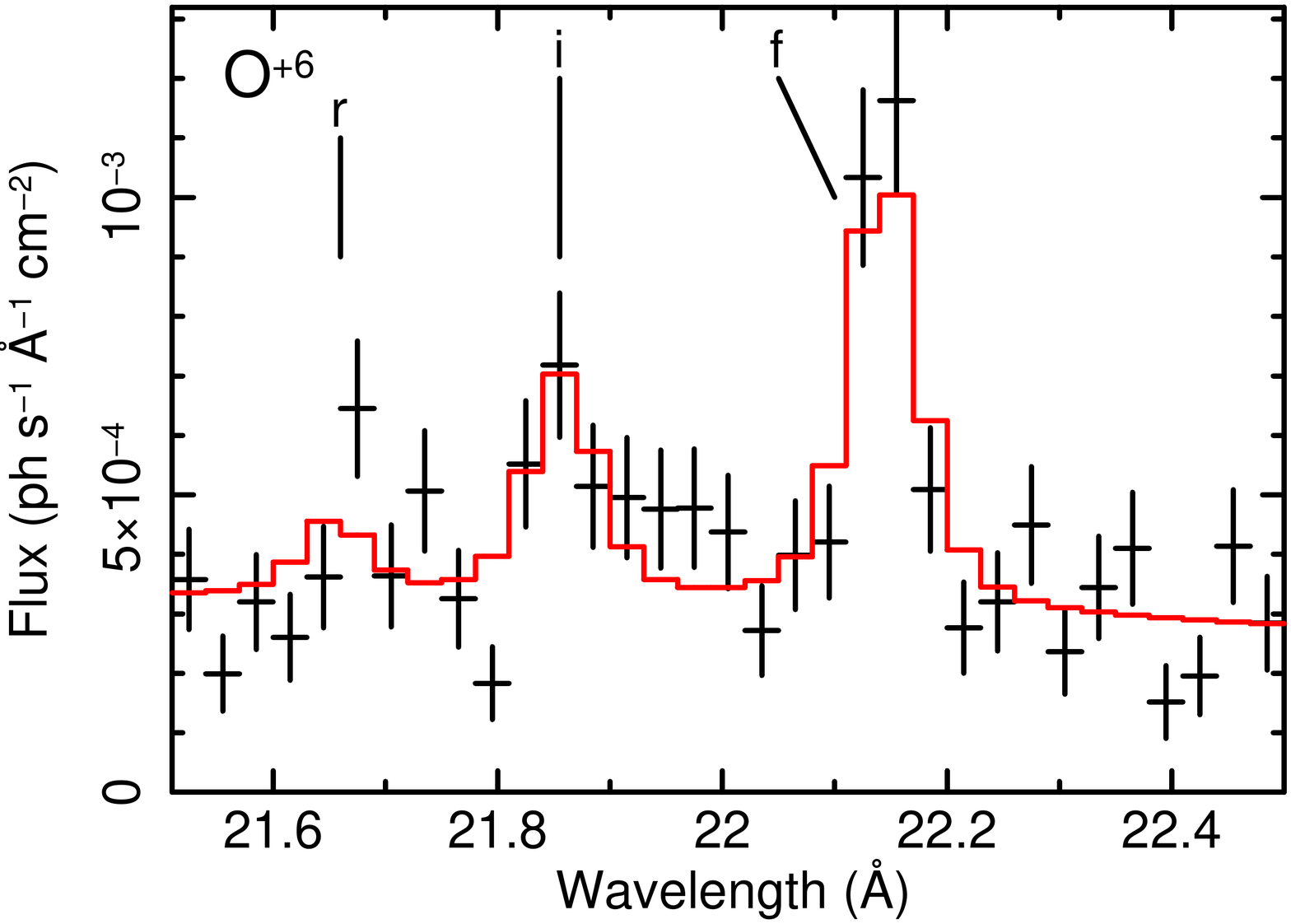}
        \caption{The two prominent He-like triplets in the 0606321901 spectrum of NGC 4051. From left to right in each are the resonance (r), intercombination (i), and forbidden (f) lines. The intercombination flux is particularly high, yielding a high density measurement. Two more features are evident; Broad residuals around 22\AA\ and heavily self-absorbed resonance lines \citep{Pounds11II}.}\label{fig:triplets}
    \end{figure}

    NGC 4051 provides bright He-like emission lines of N$^{+5}$, O$^{+6}$ and marginally Ne$^{+9}$, cleanly discerned in the low states. 
    If these are near the critical $f/i$ flux ratios \citep{Porquet0}, they allow for accurate density diagnostics. This is not possible in all observations, because both clean unabsorbed emission lines and high enough S/N is needed for accurate determination of the line ratios.
    
    The method for finding densities is detailed by \citet{Porquet0}. 
    In short, the $f/i$ ratio is sensitive to electron density, where in a critical regime, element dependent, of $10^8\lesssim n_e\lesssim 10^{12}$ cm$^{-3}$ it is an accurate estimator of $n_e$. 
    In Figure \ref{fig:triplets} we present the O$^{+6}$ and N$^{+5}$ triplets. 
    The continuum is fit locally as to not be affected by global factors, and three Gaussians are fit with their rest-frame line centers frozen, and their redshifts tied together. 
    The measurement of fluxes and uncertainties is done using the CFLUX convolution model within the HEASARC package XSPEC. The continuum is allowed to vary so the uncertainties include continuum uncertainty as well. As the width of the narrow lines is unresolved we freeze them as well to $10^{-6}$\AA, well below the RGS resolution.
    When tested, the fit is completely insensitive to this parameter. Flux measurements for all three observations and both triplets are presented in Table \ref{table:lines}. Note since CFLUX yields flux measurements in a log scale, Uncertainties on the ratios are derived directly from the exponential uncertainty.
    
    \begin{table*}
        \centering
        \caption{Fluxes, ratios, and their uncertainties are derived directly from the CFLUX model which fits $\log$(flux).}\label{table:lines}
        \begin{tabular}{c|ccccccc}
            Observation & r & i & f & f/i  & $n_e$ & $R_{n_e}$ & $R_{UV}$ \\
            &\multicolumn{3}{c}{$10^{-14}$ erg cm$^{-2}$ s$^{-1}$} &  & $10^{10}$ cm$^{-3}$ & \multicolumn{2}{c}{$10^{15}$ cm} \\             
            \hline\hline
            &\multicolumn{7}{c}{O$^{+6}$} \\ 
            \hline                           
            0606320401 & $<1.6$ & $2.7_{-2.1}^{+1.2}$ & $10_{-2.4}^{+1.6}$  & $3.8_{-2}^{+3}$     & $<5.4$              & $>4$              & $>2$\\
            0606321901 & $<2.6$ & $3.9_{-1.7}^{+1.3}$ & $7.6_{-1.9}^{+1.6}$ & $1.9_{-0.8}^{+0.9}$ & $2.9_{-1.9}^{+6.8}$ & $5_{-2}^{+3}$     & $ 2_{-1}^{+1}$\\
            0606322101 & $<2.2$ & $1.9_{-1.7}^{+1.0}$ & $11_{-2.3}^{+1.0}$  & $5.6_{-3}^{+5}$     & $<2.8$              & $>5$              & $>2$ \\
            &\multicolumn{7}{c}{N$^{+5}$} \\
            \hline
            0606320401 & $...$  & $<1.8$               & $2.4_{-1.7}^{+0.88}$ & $2.1_{-1.9}^{+2.4}$ &$0.9_{-0.7}^{+26}$ &$8_{-7}^{+12}$ & $4_{-3}^{+6}$\\
            0606321901 & $<0.3$ & $1.7_{-1.0}^{+0.68}$ & $3.1_{-1.1}^{+0.85}$ & $1.8_{-0.9}^{+1.0}$ &$1_{-0.5}^{+3.3}$  &$7_{-4}^{+4}$ & $3_{-1}^{+2}$\\
            0606322101 & $<1.5$ & $...$                & $4.4_{-1.3}^{+1.1}$  & $...$               & $...$             & $...$              & $...$              \\
        \end{tabular}                                                                                    
    \end{table*}

    \subsection{Measuring $n_e$ and $R_{n_e}$}
\begin{figure*}
        \includegraphics[width=\linewidth]{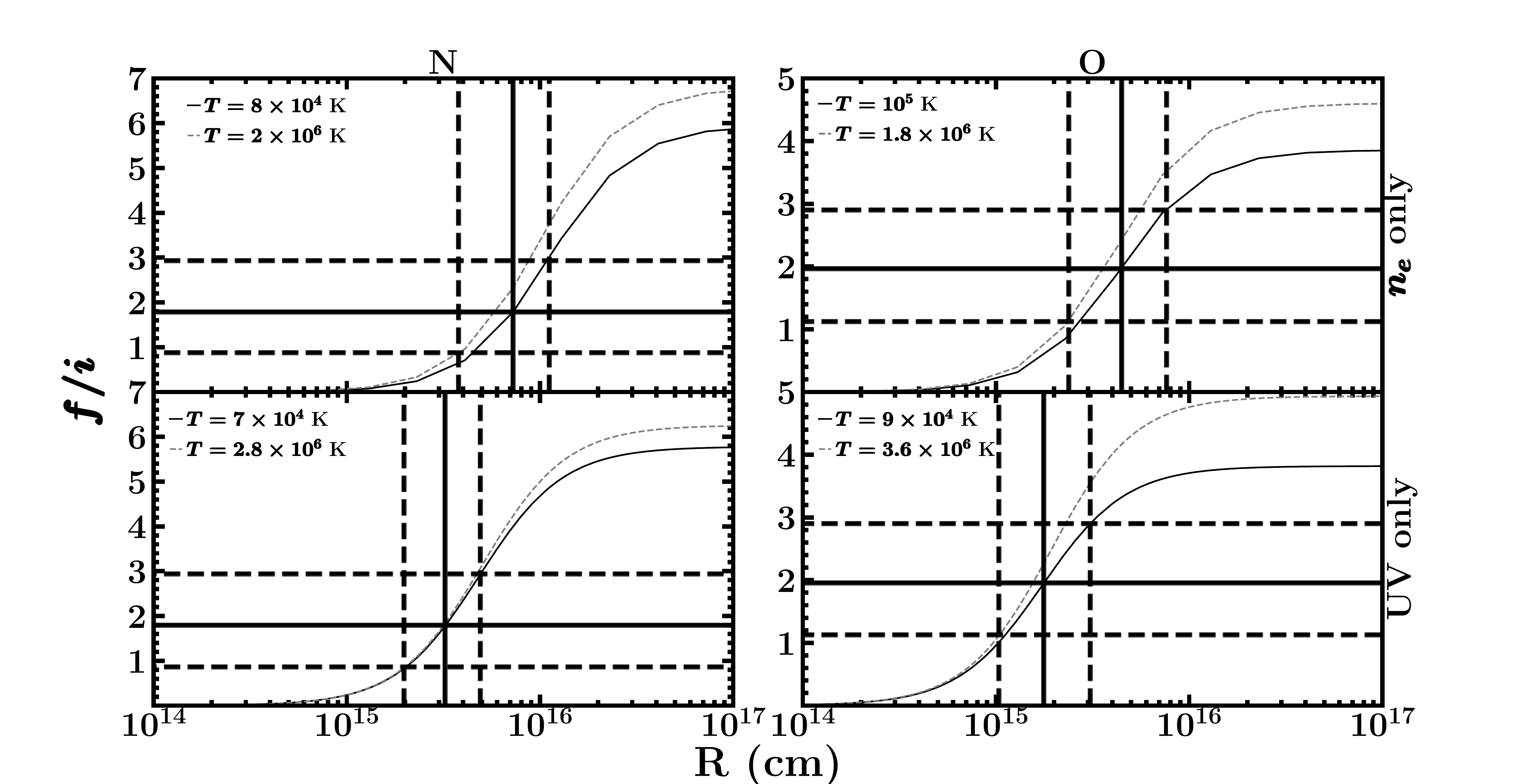}
		\caption{Distances from the continuum source estimated from the $f/i$ ratio for the 0606321901 observation, assuming intercombination enhancement either by electron collisions only (top panels) or UV pumping only (bottom panels). Distances are estimated independently using both O and N He-like emission lines, see Sec. \ref{sec:lines}. The two curves represent the extreme temperatures from \citet{Porquet0}. \textbf{Horizontal lines} show measured $f/i$ ratios, with the best fit solid and the 90\% uncertainty range dashed. \textbf{Vertical lines}: The solid line is the intersection of the best fit distance with the low $T$ curve; The dashed lines are the intersections of the uncertainty limits with the least restrictive temperatures. An estimate of $T$ may provide slightly better constraints.}\label{fig:dist9}
    \end{figure*}

    We calculate the intersections of the measured $f/i$ ratio with the theoretical graph \citep{Porquet0} to obtain $n_e$. The uncertainties are similarly derived from the $90\%$ uncertainties on the ratio. Note \citet{Porquet0} provide different curves for different temperatures. For the quoted value we use the lower temperature appropriate for photoionized plasma, and the uncertainties from the temperature yielding the most conservative uncertainty (See Fig. \ref{fig:dist9}). If $T$ is inferred in some other way these could be improved slightly.    
    
    To convert the electron densities to distances we use the ionization parameter:
    \begin{align}
    \xi=\frac L {n_e R_{n_e}^2} \Rightarrow R_{n_e}=\sqrt{\frac L {n_e \xi}}
    \end{align}
    For the ionizing luminosity $L$ we integrate the SED presented in \citet{Maitra11} to get $L=3.46\times10^{42}$~erg~s$^{-1}$, and use $\log\xi=0.82$ (cgs, erg\,cm\,s$^{-1}$) from the photoionized component in the global fit (see Section \ref{sec:global}). This allows us to convert the plots from \citet{Porquet0} from $n_e$ to $R_{n_e}$, the distance from the ionizing source. 
    Observation 0606321901 provides the most constraining measurements of distance, with both N$^{+5}$, $7^{+4}_{-4}\times10^{15}$ cm, and O$^{+6}$,  $5^{+3}_{-2}\times10^{15}$ cm, agreeing. Results are presented in Figure \ref{fig:dist9} (top panels), and see Table \ref{table:lines} for all measurements.    
    The other observations yield consistent results as well, but do not constrain distances.
    
    \subsection{$R_{UV}$}
    UV pumping can mimic the intercombination enhancement effect of electron collision. We solve the population equations assuming the intercombination levels are enhanced only through UV pumping, rather than collisions. 
    We denote the populations of the upper levels of the intercombination transitions as $n_{3,4,5}$, the population of the upper level of the forbidden line is $n_2$, and $n_1$ is that of the ground level.
    The population equations are:
    \begin{align}   
    \dot{n}_i &= n_e R_i(T)+n_2 J_{2\rightarrow i}-n_i(A_{i\rightarrow2} + A_{i\rightarrow1}),~~i=3,4,5\label{eq:ndoti} \\
    \dot{n}_2 &= n_e R_2(T)+\sum_{i=3}^5(n_i A_{i\rightarrow2} - n_2 J_{2\rightarrow i}) - n_2 A_{2\rightarrow1}\label{eq:ndot2}
    \end{align}   
    $R_i$ are the rate coefficients for radiative recombination (including cascades) detailed in Tables 4 (N$^{+5}$) and 5 (O$^{+6}$) of \citet{Porquet0}, $A_{i\rightarrow j}$ is the Einstein spontaneous decay coefficient given in Table 2 of \citet{Porquet0}. $n_e$ cancels out in the final line intensity ratios. Finally $J$ is the UV excitation rate coefficient:
    \begin{align}
    J_{2\rightarrow i}=&\int \frac {L(\nu)}{4\pi R^2} \sigma(\nu) \phi_{2\rightarrow i}(\nu) \dd\nu= \notag\\
    =&\frac {L(\nu_{2\rightarrow i})}{4\pi R^2} \sigma(\nu_{2\rightarrow i}),~~i=3,4,5
    \end{align}    
    where the line profile $\phi$ is assumed here to be a Dirac $\delta$ function centered around the $2\rightarrow i$ transition, and $\sigma$ is the relevant cross-section. The values for the wavelengths corresponding to the energy difference $2\rightarrow i$ between levels are: 1623.6\AA, 1638.3\AA, 1639.9\AA\ for O; and 1896.8\AA, 1907.3\AA, 1907.9\AA\ for N.

	The UV flux is taken from the IUE spectrum swp51045\footnote{We would like to thank Dr. Derck Massa, Dr. Nancy Oliversen, and Ms. Patricia Lawton, then members of the GSFC Astrophyics Data Facility (ADF) staff under direction of Dr. Michael Van Steenberg for providing this MAST image.}.
    While these wavelengths fall on narrow emission lines, we take only the continuum UV flux, as the UV lines are either produced at the same place as the X-ray emission lines, or much farther out. 
    The continuum flux between 1400\AA\ and 2000\AA\ is relatively flat, consequently we used $F(\nu_i-\nu_2)=F_{UV}=1.5\times10^{-14}$ erg cm$^{-2}$ s$^{-1}$ \AA$^{-1}$ for all wavelengths. Given that the spectrum was obtained during 1994, the variability of \obj, and that a factor 3 in flux results in only a factor 2 change in distance, we find this assumption reasonable.
    
    We solve the system of Eq. \ref{eq:ndoti},\ref{eq:ndot2} in steady state:
    \begin{align}
    \dot{n}_i=0
    \end{align}
    for $i=2,3,4,5$. The resulting solution provides the $f/i$ ratio as a function of distance given UV is the sole reason for intercombination enhancement. These are plotted in the lower panels of Figure \ref{fig:dist9}.
    
    This measurement, $R_{UV}=2\pm1$, implies lower distances by a factor of $\sim2$ compared with the pure collisional analysis, $R_{n_e}=5^{+3}_{-2}$, though is consistent with it and may be due to the uncertainty in $F_{UV}$. 
    The two methods should probe the same distance, as the intensity of UV pumping is directly related to the distance to the UV source, but the density diagnostic is related to this distance as well. 
    This is due to the fact we use $\xi$ to convert the density to distance, which relies on the ionizing luminosity which peaks in the UV.
	A lower density plasma, for example, to be at the same ionization level would have to be farther from the source. This would drive the line ratio towards the low density limit, but note this would be true for UV pumping as well, since less UV photons are available to pump the intercombination line.
    As such, both processes may contribute to the intercombination enhancement, and in any case we cannot discern between the two cases.    
    The distance where one process significantly contributes also sees a significant contribution from the other.
    
    \section{Global model - a photoionized emitter}\label{sec:global}
    \begin{figure*}
        \includegraphics[trim=1.5cm 0.5cm 1.5cm 0.5cm,clip,width=\linewidth]{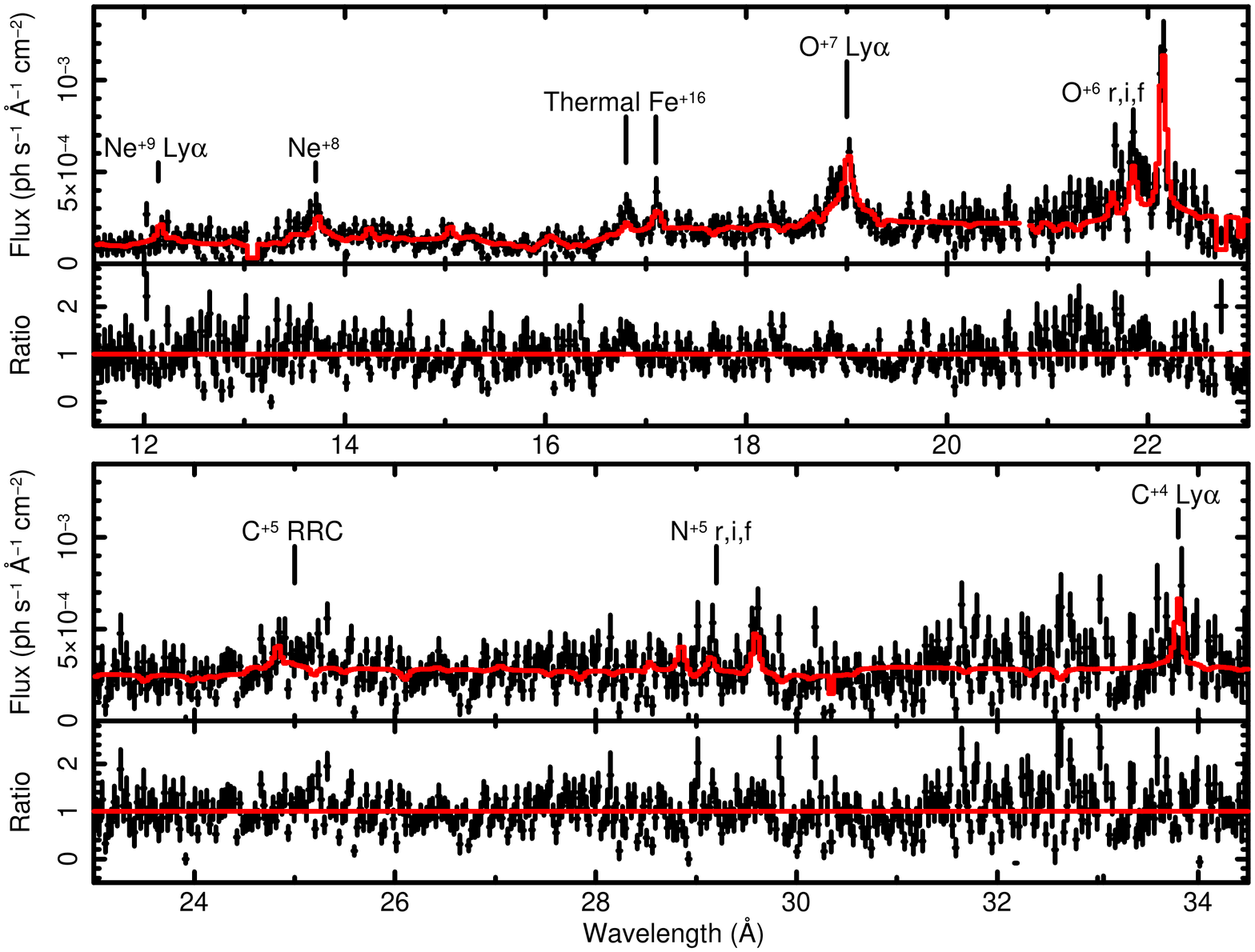}
        \caption{The data, rebinned by a factor of 3 for plotting purposes, and the best fit model in red with the data/model ratio below. The bright O$^{+7}$ lines is due to a high-$\xi$ component, both narrow and broad, the latter also seen in the O$^{+6}$ residuals. See zoom in Fig. \ref{fig:triplets}. The bright He-like O$^{+6}$ lines and its edge at 16.8\AA\ are due to a lower $\xi$. The Fe$^{+16}$ lines around 15\AA\ and 17\AA\ are clear indicators of a thermal component.}\label{fig:album}
    \end{figure*}

    We model the full spectrum of observation 0606321901, which best constrains the distances. 
    The data and global model are shown in Figure \ref{fig:album}.
    The fit was done on the 8\AA-37\AA\ band, but featureless edges of the spectrum are clipped.
    Since absorption features are hard to discern in the low state, we follow the previous works listed in Section \ref{sec:prev}, and examine the high state. We find the only prominent absorption is due to an $10^4$~km~s$^{-1}$ outflow. This velocity is accurately determined in the high state, also reported by \citet{Pounds11I}. This is the only absorption component clearly detected in the low state, particularly due to the O$^{+6}$ edge at 16.8\AA .
    
    The Fe$^{+16}$ prominent emission lines around 15 and 17\AA\ are a clear indication of a thermal plasma of $kT=0.7$ keV, as the rates of recombination to L-shells in a photoionized plasma are too low. At $\log\xi\approx2$ where the Fe$^{+16}$ abundance peaks, the predicted flux of these lines by the modeled photoionized plasma are still almost 10 times less. At the best-fitted $\log\xi=0.8$ value of the photoionized component, these lines are completely non-existent.
    
    \begin{figure}
        \centering
        \includegraphics[trim=0.5cm 0.5 2cm 0.8cm,clip,width=\linewidth]{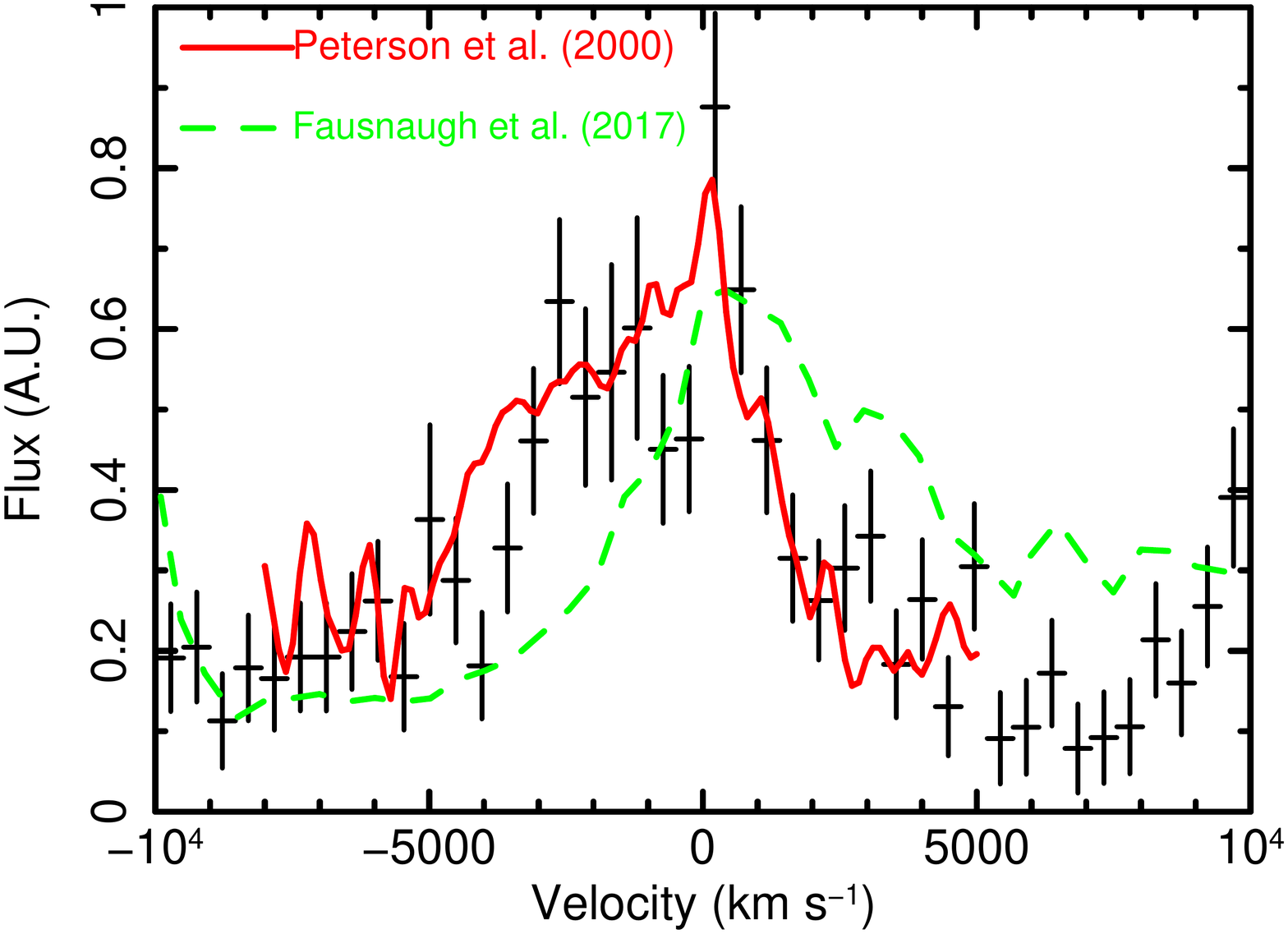}
        \caption{The O$^{+7}$ line compared to the RMS He$^{+1}\,\lambda4686$\,\AA\ line profile. 
            Fluxes in erg cm$^{-2}$ s$^{-1}$ \AA$^{-1}$ were scaled and shifted for the comparison.
            The similarity to the line measured by \citet{Peterson00} is striking. The broadening in the line observed by \citet{Fausnaugh17} is similar as well, though the line is shifted. 
        The similar profiles indicate a similar origin in the BLR.}\label{fig:oviii}
    \end{figure}
    
    The best fit model continuum is comprised of a powerlaw and a blackbody, as a powerlaw cannot fit the spectrum well which has been previously noted by e.g. \citet{Collinge01}. This is a phenomenological description of the continuum, and we do not derive physical parameters from it.
	Next, the model includes a photoionized component which produces both emission and absorption features, a highly photoionized component which produces both narrow and broad ($v_\mathrm{turb}=\sqrt2\sigma\sim4000$ km\,s$^{-1}$) emission, and a thermal emission component. The best fitted model parameters and 90\% uncertainties are given in Table~\ref{table:global}.	
	The emitting components' redshifts are frozen to that of \obj, $z=0.00216$. The population file used for the absorption and emission photoionized models in XSPEC\footnote{https://heasarc.gsfc.nasa.gov/docs/xanadu/xspec/} within the HEASoft package\footnote{https://heasarc.gsfc.nasa.gov/lheasoft/}, WARMABS and PHOTEMIS\footnote{https://heasarc.gsfc.nasa.gov/xstar/xstar.html by Tim Kallman} respectively, are for $n_e=10^{10}$ cm$^{-3}$ as found in Section \ref{sec:lines}.
    
    The lower $\log\xi=0.8$ photoionized component fits most of the emission lines, and in particular the bright He-like ones. The same $\xi$ also produces the $10^4$\,km\,s$^{-1}$ absorbing outflow, and especially the O$^{+6}$ edge. Another component is the more ionized $\log\xi=1.4$, producing the narrow O$^{+7}$ and the Ne$^{+9}$ lines. This $\xi$ also produces a broad emission component, most prominent in the broad shoulder of the O$^{+7}$ Ly$\alpha$ line.
    Interestingly, the Emission Measure (EM) of the $\log\xi=1.4$ broad component, EM$=\int n_H n_e \dd V=2\times10^{63}$ cm$^{-3}$, matches that of the $\log\xi=0.8$ component which produces the He-like lines that provide the high density line diagnostics. 
    In Section \ref{sec:blr} we use this to estimate the size of the BLR.
      
    In Figure \ref{fig:oviii} we compare the O$^{+7}$ line profile to the RMS He$^{+1}\,\lambda4686$\,\AA\ profiles taken from \citet{Peterson00} and \citet{Fausnaugh17}. 
    The similarity with these two profiles and their blue wing is remarkable, in particular that of \citet{Peterson00}, suggesting the two ions originate from the same region, or at least share kinematics. This reinforces the evidence for an X-ray component in a stratified BLR. The mean C$^{+3}$ profile from \citet{Peterson00} is similarly broad, though it is heavily absorbed, as seen in \citet{Kramer12}.
          
	\begin{table*}
		\centering
		\caption{Physical model parameters}\label{table:global}
		\begin{tabular}{l|lll}
			Component      & Parameter         & Units                            &   Value                 \\
			\hline
			\textbf{Continuum} &&&\\			 
			Powerlaw       & $\Gamma$  &                                  & $2.9^{+0.06}_{-0.06}$    \\
			               & norm      & ph keV$^{-1}$ cm$^{-2}$ s$^{-1}$ & $1.8^{+0.1}_{-0.1}\times10^{-3}$ \\
			Blackbody      & kT        & keV                              & $0.1^{+0.003}_{-0.007}$               \\
			               & Luminosity& erg s$^{-1}$                     & $6.4^{+0.4}_{-0.8}\times10^{41}$ \\
			\hline
			\textbf{Highly } &&&\\
			\textbf{photoionized} & $\log\xi$  & [erg cm s$^{-1}$]                & $1.4^{+0.2}_{-0.03}$\\
			Broad emission & $v_\mathrm{turb}$ & km s$^{-1}$                      & 3800$^{+900}_{-800}$ \\ 
			               & EM                & cm$^{-3}$                        & $2_{-0.4}^{+0.4}\times10^{63}$\\ 
			Narrow emission& $v_\mathrm{turb}$ & km s$^{-1}$                      & unresolved \\ 
						   & EM                & cm$^{-3}$                        & $2_{-0.5}^{+0.6}\times10^{62}$\\ 
			\hline
			\textbf{Photoionized} &&&\\
		                   & $\log\xi$          & [erg cm s$^{-1}$]                & $0.78^{+0.06}_{-0.01}$\\
						   & $N/N_\odot$        & \citet{Asplund09}                & $2.5^{+0.6}_{-0.3}$   \\
			Absorption 	   & column             & cm$^{-2}$                        & $6.4^{+0.6}_{-0.7}\times10^{21}$                \\
			               & Velocity           & km s$^{-1}$                      & $-10000_{-70}^{+135}$ \\

		    Emission       & EM                 & cm$^{-3}$                        & $2_{-0.4}^{+1.6}\times10^{63}$      \\ 
             & Velocity & km s$^{-1}$ & 0 (fixed) \\
			\hline                                                                
            \textbf{Thermal}(CIE) &&&\\
                           & kT                & keV                              & $0.74^{+0.07}_{-0.04}$ \\
                           & EM                & cm$^{-3}$                        & $3_{-0.63}^{+0.65}\times10^{62}$   \\ 
		\end{tabular}
	\end{table*}
     
    \section{The X-ray BLR}\label{sec:blr}
    Line profiles of the \obj\ BLR were analyzed in the optical and UV. 
    From the optical He$^{+1}$ line in the RMS spectrum, \citet{Peterson00} find a  FWHM velocity of $\sim5000$ km s$^{-1}$, with a blue-ward asymmetry of $\sim1400$ km s$^{-1}$ indicating an out-flowing wind. Similarly, in the UV they find asymmetric and broad profiles in the He$^{+1}$ and C$^{+3}$ lines.
    \citet{Shemmer03} find the distance to the BLR to be approximately $8\pm4\times10^{15}$ cm from reverberation mapping, and recently \citet{Fausnaugh17} improved uncertainties and estimated $5.8\pm0.9\times10^{15}$ cm. 

    Previous X-ray measurements place several of the outflow components at distances consistent with the UV and optical measurements of the BLR, with modeling indicating an X-ray BLR, e.g., by \citet{Costantini07}. 
    In the present work we find direct evidence of $\sim10^{10}$ cm$^{-3}$ plasma consistent with BLR densities. The plasma is emitting He-like lines which indicates a distance of $10^{15}\lesssim R_\mathrm{BLR}\lesssim10^{16}$ cm, similar to that of \citet{Fausnaugh17}. We also measure a $v_\mathrm{turb}=3800^{-800}_{+900}$ km\,s$^{-1}$ wide profile in the O$^{+7}$ line, slightly broader than the He$^{+1}\lambda4686$\,\AA\ RMS profile where $v_\mathrm{turb}=2400_{-54}^{51}$ km\,s$^{-1}$ \citep{Fausnaugh17}. This suggests the X-ray BLR is slightly closer to the source despite the consistent distance measurements. The EM and resulting mass measurement below indicates the X-ray emitting plasma occupies a small fraction of the BLR, likely the inner most part.
    
    If the BLR is radiatively compressed as in the RPC model \citep{Baskin14}, such that the lower ionization gas has higher density and all ionization states originate at the same distance, the UV and optical BLR should be denser than the present density measurements. 
    For example,C$^{+3}$ peaks at $\log\xi=-0.6$, 25 times lower than the our $\log\xi=0.8$ component. If they originate at the same distance, then the UV plasma is 25 times denser than the X-ray plasma, that is $n_e=8\times10^{11}$ cm$^{-3}$.
    The H$\beta$ distance similar to ours \citep{Shemmer03,Fausnaugh17} would imply the optical plasma is at least an additional order of magnitude denser.
             	
    Using the common EM of the broad component and the high density component $EM=2\times10^{63}$ cm$^{-3}$ (Table \ref{table:global}) associated with the BLR, and the density measured in the O lines for observation 0606321901, $3\times10^{10}$ cm$^{-3}$ (Table \ref{table:lines}) we can estimate the volume of the BLR as $EM/n_e^2$. Assuming the X-ray BLR is an inner shell of the BLR, its scale length, $l_\mathrm{BLR}$ is:
    \begin{align}
    l_\mathrm{BLR}=\frac 1 {\Omega R_\mathrm{BLR}^2} \frac{\mathrm{EM}}{n_e^2}=1.3_{-0.2}^{+0.9}\times10^{11}\,\mathrm{cm}
    \end{align}
    where we take the opening angle of the X-ray BLR to be $\Omega=1.2\pi$ and $R_\mathrm{BLR}=5\times10^{15}$ cm.
    This scale length stretches $\sim(1.3\times10^{11}/5\times10^{15})=3\times10^{-5}$, indicating this is a thin shell. This scale length along with the same density yields a column density of $N=n_el_\mathrm{BLR}=4\times10^{21}$ cm$^{-2}$, consistent with the X-ray BLR being a thin absorption shell.
    
    We can further employ our measurements to estimate the total mass of the X-ray BLR:
    \begin{align}
    M_\mathrm{BLR}=m_p EM/n_e\approx5\times10^{-5}\ M_\odot
    \end{align}
    where $m_p$ is the proton mass.
    The X-ray BLR is not a very massive component, only a tiny fraction of the black hole mass, $\sim10^6\ M_\odot$.
    This low mass again indicates we are observing the thin inner shell of the BLR, as it is orders of magnitude less than previous mass estimates.
    The dust models of \citet{Czerny15} for example estimate a Seyfert $M_\mathrm{BLR}$ of $\sim10$ M$_\odot$.
    
	From all these we associate both the broad and low ionization emission with the BLR. This means ionization ranges from $\log\xi=-2$ in the optical to $\log\xi=1.5$. A stratified BLR such as proposed by \citet{Costantini07} is possible, but the matching EM of the broad, highly ionized component and the lower ionization component suggest at least these two are co-spatial. 
    
    The high density along with the small mass indicate that the X-ray BLR could possibly be photo-driven off the accretion disk \citep{Proga00}, though other mechanisms such as MHD winds \citep{Fukumura10} are not ruled out.	
    This makes ionized outflows observed at these distances a method for the AGN system to supply mass to the BLR. 

    \section{Conclusion}
    We measure for the first time directly the density, distance, size, and mass of the BLR in X-rays. The distance and density are found to be in-line with those measured in the optical and UV wavelengths of the BLR, also matching kinematics. 
    On the other hand, the mass estimate of the X-ray BLR is orders of magnitude less than those of the UV-optical BLR, indicating we may be seeing only a fraction, the inner part, of the BLR.
    To sum, the currently measured X-ray BLR parameters:
    
    \begin{tabular}{l@{ : }l@{ $\sim$ }l}
    	--Distance    & $R_\mathrm{BLR}$ & $5\times10^{15}$ cm. \\
    	--Density     & $n_e$            & $3\times10^{10}$ cm$^{-3}$. \\
    	--Scale length& $l_\mathrm{BLR}$ & $10^{11}$ cm. \\
    	--Mass        & $M_\mathrm{BLR}$ & $5\times10^{-5}$ M$_\odot.$ \\
    	--Ionization  & \multicolumn{2}{c}{$-2<\log\xi<1.5$ (optical to X-rays).} \\
    \end{tabular}
    
     In terms of predicting mass ejection mechanisms from the accretion disk and the near-BH environment, considering the entire range from the event horizon at $3\times10^{11}$ cm to the BLR at $5\times10^{15}$ cm, for \obj, may be important to better understand the feedback of the AGN system with the host galaxy.
    
    \acknowledgments
    	We thank Ari Laor for insightful discussions on the UV and optical properties of the BLR. We also thank Tim Kallman for all his support with XSTAR and WARMABS/PHOTEMIS.
	
    \bibliographystyle{apj}
    \bibliography{../agn,../../general,NGC4051}

\begin{thebibliography}{31}
\expandafter\ifx\csname natexlab\endcsname\relax\def\natexlab#1{#1}\fi

\bibitem[{{Asplund} {et~al.}(2009){Asplund}, {Grevesse}, {Sauval}, \&
  {Scott}}]{Asplund09}
{Asplund}, M., {Grevesse}, N., {Sauval}, A.~J., \& {Scott}, P. 2009, \araa, 47,
  481

\bibitem[{{Baldwin} {et~al.}(1995){Baldwin}, {Ferland}, {Korista}, \&
  {Verner}}]{Badlwin95}
{Baldwin}, J., {Ferland}, G., {Korista}, K., \& {Verner}, D. 1995, \apjl, 455,
  L119

\bibitem[{{Baskin} {et~al.}(2014){Baskin}, {Laor}, \& {Stern}}]{Baskin14}
{Baskin}, A., {Laor}, A., \& {Stern}, J. 2014, \mnras, 445, 3025

\bibitem[{{Blustin} \& {Fabian}(2009)}]{Blustin09}
{Blustin}, A.~J. \& {Fabian}, A.~C. 2009, \mnras, 399, L169

\bibitem[{{Collinge} {et~al.}(2001){Collinge}, {Brandt}, {Kaspi}, {Crenshaw},
  {Elvis}, {Kraemer}, {Reynolds}, {Sambruna}, \& {Wills}}]{Collinge01}
{Collinge}, M.~J., {Brandt}, W.~N., {Kaspi}, S., {et~al.} 2001, \apj, 557, 2

\bibitem[{{Costantini} {et~al.}(2007){Costantini}, {Kaastra}, {Arav}, {Kriss},
  {Steenbrugge}, {Gabel}, {Verbunt}, {Behar}, {Gaskell}, {Korista}, {Proga},
  {Quijano}, {Scott}, {Klimek}, \& {Hedrick}}]{Costantini07}
{Costantini}, E., {Kaastra}, J.~S., {Arav}, N., {et~al.} 2007, \aap, 461, 121

\bibitem[{{Costantini} {et~al.}(2016){Costantini}, {Kriss}, {Kaastra},
  {Bianchi}, {Branduardi-Raymont}, {Cappi}, {De Marco}, {Ebrero}, {Mehdipour},
  {Petrucci}, {Paltani}, {Ponti}, {Steenbrugge}, \& {Arav}}]{Costantini16}
{Costantini}, E., {Kriss}, G., {Kaastra}, J.~S., {et~al.} 2016, \aap, 595, A106

\bibitem[{{Czerny} {et~al.}(2015){Czerny}, {Modzelewska}, {Petrogalli}, {Pych},
  {Adhikari}, {{\.Z}ycki}, {Hryniewicz}, {Krupa}, {{\'S}wie{\c t}o{\'n}}, \&
  {Niko{\l}ajuk}}]{Czerny15}
{Czerny}, B., {Modzelewska}, J., {Petrogalli}, F., {et~al.} 2015, Advances in
  Space Research, 55, 1806

\bibitem[{{Fausnaugh} {et~al.}(2017){Fausnaugh}, {Grier}, {Bentz}, {Denney},
  {De Rosa}, {Peterson}, {Kochanek}, {Pogge}, {Adams}, {Barth}, {Beatty},
  {Bhattacharjee}, {Borman}, {Boroson}, {Bottorff}, {Brown}, {Brown},
  {Brotherton}, {Coker}, {Crawford}, {Croxall}, {Eftekharzadeh}, {Eracleous},
  {Joner}, {Henderson}, {Holoien}, {Horne}, {Hutchison}, {Kaspi}, {Kim},
  {King}, {Li}, {Lochhaas}, {Ma}, {MacInnis}, {Manne-Nicholas}, {Mason},
  {Montuori}, {Mosquera}, {Mudd}, {Musso}, {Nazarov}, {Nguyen}, {Okhmat},
  {Onken}, {Ou-Yang}, {Pancoast}, {Pei}, {Penny}, {Poleski}, {Rafter},
  {Romero-Colmenero}, {Runnoe}, {Sand}, {Schimoia}, {Sergeev}, {Shappee},
  {Simonian}, {Somers}, {Spencer}, {Starkey}, {Stevens}, {Tayar}, {Treu},
  {Valenti}, {Van Saders}, {Villanueva}, {Villforth}, {Weiss}, {Winkler}, \&
  {Zhu}}]{Fausnaugh17}
{Fausnaugh}, M.~M., {Grier}, C.~J., {Bentz}, M.~C., {et~al.} 2017, \apj, 840,
  97

\bibitem[{{Fukumura} {et~al.}(2010){Fukumura}, {Kazanas}, {Contopoulos}, \&
  {Behar}}]{Fukumura10}
{Fukumura}, K., {Kazanas}, D., {Contopoulos}, I., \& {Behar}, E. 2010, \apj,
  715, 636

\bibitem[{{Kaastra} {et~al.}(2002){Kaastra}, {Steenbrugge}, {Raassen}, {van der
  Meer}, {Brinkman}, {Liedahl}, {Behar}, \& {de Rosa}}]{Kaastra02}
{Kaastra}, J.~S., {Steenbrugge}, K.~C., {Raassen}, A.~J.~J., {et~al.} 2002,
  \aap, 386, 427

\bibitem[{{Kraemer} {et~al.}(2012){Kraemer}, {Crenshaw}, {Dunn}, {Turner},
  {Lobban}, {Miller}, {Reeves}, {Fischer}, \& {Braito}}]{Kramer12}
{Kraemer}, S.~B., {Crenshaw}, D.~M., {Dunn}, J.~P., {et~al.} 2012, \apj, 751,
  84

\bibitem[{{Lobban} {et~al.}(2011){Lobban}, {Reeves}, {Miller}, {Turner},
  {Braito}, {Kraemer}, \& {Crenshaw}}]{Lobban11}
{Lobban}, A.~P., {Reeves}, J.~N., {Miller}, L., {et~al.} 2011, \mnras, 414,
  1965

\bibitem[{{Maitra} {et~al.}(2011){Maitra}, {Miller}, {Markoff}, \&
  {King}}]{Maitra11}
{Maitra}, D., {Miller}, J.~M., {Markoff}, S., \& {King}, A. 2011, \apj, 735,
  107

\bibitem[{{Mason} {et~al.}(2004){Mason}, {Branduardi-Raymont}, {Ogle}, {Page},
  {Puchnarewicz}, \& {Salvi}}]{Mason04}
{Mason}, K.~O., {Branduardi-Raymont}, G., {Ogle}, P.~M., {et~al.} 2004,
  Advances in Space Research, 34, 2610

\bibitem[{{Miller} {et~al.}(2018){Miller}, {Cackett}, {Zoghbi}, {Barret},
  {Behar}, {Brenneman}, {Fabian}, {Kaastra}, {Lohfink}, {Mushotzky}, {Nandra},
  \& {Raymond}}]{Miller18}
{Miller}, J.~M., {Cackett}, E., {Zoghbi}, A., {et~al.} 2018, \apj, 865, 97

\bibitem[{{Ogle} {et~al.}(2004){Ogle}, {Mason}, {Page}, {Salvi}, {Cordova},
  {McHardy}, \& {Priedhorsky}}]{Ogle04}
{Ogle}, P.~M., {Mason}, K.~O., {Page}, M.~J., {et~al.} 2004, \apj, 606, 151

\bibitem[{{Peterson} {et~al.}(2000){Peterson}, {McHardy}, {Wilkes}, {Berlind},
  {Bertram}, {Calkins}, {Collier}, {Huchra}, {Mathur}, {Papadakis}, {Peters},
  {Pogge}, {Romano}, {Tokarz}, {Uttley}, {Vestergaard}, \&
  {Wagner}}]{Peterson00}
{Peterson}, B.~M., {McHardy}, I.~M., {Wilkes}, B.~J., {et~al.} 2000, \apj, 542,
  161

\bibitem[{{Porquet} \& {Dubau}(2000)}]{Porquet0}
{Porquet}, D. \& {Dubau}, J. 2000, \aaps, 143, 495

\bibitem[{{Pounds} {et~al.}(2004){Pounds}, {Reeves}, {King}, \&
  {Page}}]{Pounds04}
{Pounds}, K.~A., {Reeves}, J.~N., {King}, A.~R., \& {Page}, K.~L. 2004, \mnras,
  350, 10

\bibitem[{{Pounds} \& {Vaughan}(2011{\natexlab{a}})}]{Pounds11I}
{Pounds}, K.~A. \& {Vaughan}, S. 2011{\natexlab{a}}, \mnras, 413, 1251

\bibitem[{{Pounds} \& {Vaughan}(2011{\natexlab{b}})}]{Pounds11II}
{Pounds}, K.~A. \& {Vaughan}, S. 2011{\natexlab{b}}, \mnras, 415, 2379

\bibitem[{{Proga} {et~al.}(2000){Proga}, {Stone}, \& {Kallman}}]{Proga00}
{Proga}, D., {Stone}, J.~M., \& {Kallman}, T.~R. 2000, \apj, 543, 686

\bibitem[{{Reichert} {et~al.}(1986){Reichert}, {Mushotzky}, \&
  {Holt}}]{Reichert86}
{Reichert}, G.~A., {Mushotzky}, R.~F., \& {Holt}, S.~S. 1986, \apj, 303, 87

\bibitem[{{Risaliti} {et~al.}(2011{\natexlab{a}}){Risaliti}, {Nardini},
  {Elvis}, {Brenneman}, \& {Salvati}}]{Risaliti11a}
{Risaliti}, G., {Nardini}, E., {Elvis}, M., {Brenneman}, L., \& {Salvati}, M.
  2011{\natexlab{a}}, \mnras, 417, 178

\bibitem[{{Risaliti} {et~al.}(2011{\natexlab{b}}){Risaliti}, {Nardini},
  {Salvati}, {Elvis}, {Fabbiano}, {Maiolino}, {Pietrini}, \&
  {Torricelli-Ciamponi}}]{Risaliti11b}
{Risaliti}, G., {Nardini}, E., {Salvati}, M., {et~al.} 2011{\natexlab{b}},
  \mnras, 410, 1027

\bibitem[{{Shemmer} {et~al.}(2003){Shemmer}, {Uttley}, {Netzer}, \&
  {McHardy}}]{Shemmer03}
{Shemmer}, O., {Uttley}, P., {Netzer}, H., \& {McHardy}, I.~M. 2003, \mnras,
  343, 1341

\bibitem[{{Silva} {et~al.}(2016){Silva}, {Uttley}, \& {Costantini}}]{Silva16}
{Silva}, C.~V., {Uttley}, P., \& {Costantini}, E. 2016, \aap, 596, A79

\bibitem[{{Steenbrugge} {et~al.}(2009){Steenbrugge}, {Fenov{\v c}{\'{\i}}k},
  {Kaastra}, {Costantini}, \& {Verbunt}}]{Steenbrugge09}
{Steenbrugge}, K.~C., {Fenov{\v c}{\'{\i}}k}, M., {Kaastra}, J.~S.,
  {Costantini}, E., \& {Verbunt}, F. 2009, \aap, 496, 107

\bibitem[{{Svoboda} {et~al.}(2015){Svoboda}, {Beuchert}, {Guainazzi},
  {Longinotti}, {Piconcelli}, \& {Wilms}}]{Svoboda15}
{Svoboda}, J., {Beuchert}, T., {Guainazzi}, M., {et~al.} 2015, \aap, 578, A96

\bibitem[{{Uttley} {et~al.}(2004){Uttley}, {Taylor}, {McHardy}, {Page},
  {Mason}, {Lamer}, \& {Fruscione}}]{Uttley04}
{Uttley}, P., {Taylor}, R.~D., {McHardy}, I.~M., {et~al.} 2004, Nuclear Physics
  B Proceedings Supplements, 132, 240

\end{thebibliography}

\end{document}